\documentclass[12pt,a4paper]{article}

\usepackage[a4paper,
            top=1in,
            bottom=1in,
            left=1in,
            right=1in]{geometry}

\usepackage{amsmath}
\usepackage{amssymb}
\usepackage{amsfonts}

\usepackage{graphicx}
\usepackage{epstopdf}

\usepackage{array}

\usepackage{url}
\usepackage[
    colorlinks=true,
    linkcolor=blue,
    citecolor=blue,
    urlcolor=blue
]{hyperref}

\usepackage{xcolor}
\usepackage{authblk}
\begin{document}

\title{Horizon energy fluctuations beyond Einstein's gravity: Gauss-Bonnet, Lovelock and Quantum deformed frameworks}

\author{
P.~B.~Krishna$^{1}$, Lini Devassy$^{2}$, and Titus K.~Mathew$^{3}$\\
\small
\texttt{krishnapb99@gmail.com, linidevassy@bharatamatacollege.in, titus@cusat.ac.in}\\
\small
$^{1}$Department of Space Science, St.~Albert's College,
Kochi--682018, Kerala, India\\
$^{2}$Department of Physics, Bharata Mata College,
Thrikkakara--682021, Kerala, India\\
$^{3}$Department of Physics, Cochin University of Science and Technology,
Kochi--682022, Kerala, India
}
\date{}

\maketitle
\begin{abstract}
We study thermal energy fluctuations of cosmological horizons within the canonical ensemble framework,
treating the horizon as a thermodynamic system characterized by temperature and entropy. The analysis
is performed for a class of gravitational theories, including $(n+1)$-dimensional Einstein gravity,
Gauss-Bonnet gravity, Lovelock gravity, and models incorporating quantum-deformed entropy corrections. We find that horizon energy fluctuations stabilize to a constant value in the asymptotic
de Sitter limit $\omega \to -1$, independent of the underlying gravity theory and robust against
higher-curvature and quantum corrections. It is worth mentioning that in its final de Sitter state, the universe obeys holographic equipartition condition $N_{\text{surf}} = N_{\text{bulk}}$ and in consequence the horizon entropy attains a maximum constant value, just like an ordinary macroscopic system. These findings establish a unified thermodynamic picture in which entropy maximization, holographic
equipartition, and the suppression of energy fluctuations collectively characterize the asymptotic
de Sitter universe as the equilibrium end state of cosmic evolution. 
\end{abstract}

\section{Introduction}
The intriguing connection between gravitation and thermodynamics has become a cornerstone in the quest to understand the microscopic nature of spacetime. Early insights from black hole physics revealed that event horizons possess temperature and entropy, with the entropy proportional to the horizon area and the temperature determined by surface gravity \cite{Bekenstein1973,Hawking1975,GibbonsHawking1977,Unruh1976,Bousso2002}. These discoveries strongly suggested that gravitational dynamics may admit a thermodynamic or statistical interpretation rather than being fundamentally geometric. This idea was placed on a firm conceptual footing by Jacobson, who demonstrated that Einstein’s field equations can be derived from the Clausius relation applied to local Rindler horizons, thereby identifying them as equations of state \cite{Jacobson1995,CaiKim2005,AkbarCai2007,GongWang2007,CaiCao2007}. Subsequent developments have shown that this thermodynamic perspective extends naturally beyond Einstein gravity. In particular, higher-curvature theories such as Gauss–Bonnet and Lovelock gravities modify the gravitational action through additional curvature invariants, leading to nontrivial corrections to horizon entropy and thermodynamic relations \cite{Wald1993,Cai2002,Sheykhi2010,
Brustein2009}. 

Subsequent investigations have considerably strengthened the viewpoint that gravitational dynamics are deeply rooted in thermodynamic principles, revealing gravity as an emergent phenomenon rather than a fundamental interaction \cite{Padmanabhan2010,Padmanabhan2017,Sindoni2012,Padmanabhan2015}. Within this emergent framework, Verlinde demonstrated that Newtonian gravity can be interpreted as an entropic force \cite{Verlinde2011}, while Padmanabhan independently derived gravitational dynamics by invoking the equipartition of energy among horizon degrees of freedom together with the thermodynamic relation $S=E/2T$, where $E$
represents the effective gravitational energy,$S$ 
the horizon entropy, and 
$T$ the associated temperature \cite{PadmanabhanEquip2010}.

Later, Padmanabhan proposed that the expansion of the universe may be viewed as an emergent process driven by the difference between surface and bulk degrees of freedom, a principle known as holographic equipartition \cite{Padmanabhan2010,Padmanabhan2012}. This framework has been successfully applied to reproduce standard cosmological dynamics and to provide insights into the late-time acceleration of the universe \cite{Komatsu2022}. From a thermodynamic perspective, several studies have argued that the universe evolves toward a state of maximum entropy, with the de Sitter phase representing the thermodynamically favored end state of cosmic evolution \cite{Pavon2009,Myung2008}.

An essential feature of any thermodynamic system is the presence of energy fluctuations arising from its underlying microscopic degrees of freedom. Given the thermodynamic nature of spacetime horizons, it is natural to expect that horizon energy should also exhibit statistical fluctuations. While thermal fluctuations of matter and volume energy within cosmological horizons have been widely studied—particularly in connection with primordial perturbations and structure formation—the evolution of fluctuations associated with the horizon energy itself has received comparatively little attention. Recent investigations have begun to address this issue in specific contexts, such as de Sitter space, highlighting the relevance of horizon thermodynamics for late-time cosmology \cite{Komatsu2022}.

In this work, we investigate thermal energy fluctuations of cosmological horizons within the canonical ensemble framework, with particular emphasis on gravity models extending beyond general relativity. 
 This approach provides a robust foundation for investigating thermodynamic behavior in Einstein, Gauss-Bonnet and Lovelock gravity. We assess the thermodynamic stability of these theories by examining energy fluctuations and their dependence on the equation of state parameter $\omega$. We also extend our analysis to the framework of quantum-deformed entropy, where corrections to the classical entropy-area relation are expected to influence the underlying thermodynamic structure. Motivated by recent developments in the emergent gravity paradigm, particularly the connection between the emergence of cosmic space and horizon thermodynamics, we explore how quantum corrections modify the thermodynamic stability criteria and contribute to our understanding of space time dynamics. To quantify energy fluctuations, we will find the second derivative of the entropy with respect to the inverse temperature $\beta$, which is a well-established method in canonical ensemble framework to evaluate thermal stability \cite{Namboothiri2026, Krishna2026}. 
 
The paper is organized as follows. In section II, We outline the general thermodynamic considerations relevant to energy fluctuations in canonical ensembles, establishing the statistical basis for fluctuation phenomena in macroscopic systems. In section III, we extend this framework to gravitational settings by analyzing fluctuations in horizon energy, treating spacetime horizons as thermodynamic systems endowed with temperature and entropy. The formalism is applied successively to $(n+1)$-dimensional Einstein gravity, Gauss–Bonnet gravity, and Lovelock gravity, where higher-curvature corrections modify the horizon entropy and, consequently, the associated fluctuation behavior. We further investigate horizon energy fluctuations in the case of quantum-deformed entropy, highlighting the role of microscopic corrections to the entropy–area relation and their impact on horizon thermodynamics. Building on these results, we explore the connection between horizon energy fluctuations, entropy maximization, and holographic equipartition, examining how statistical fluctuations may influence the emergence of cosmic space in section IV. We present our conclusions in section V.

\section{The thermodynamic energy fluctuations }

The basic framework describing thermodynamic fluctuations can be presented as follows.
A system that is in thermal equilibrium with a surrounding heat reservoir can be studied using the canonical ensemble approach, for which the partition function is defined as

\begin{equation}
	Z(\beta) = \int_{0}^{\infty} dE \, \exp(-\beta E)\rho(E),
\end{equation}
with $Z(\beta)$, the canonical partition function, $k_B$ the Boltzmann constant, $\beta$ connected to the temperature of the system by $1/k_B T$ and $\rho(E)$ is the density of energy states. The mean (or expected) energy of the system can be obtained from the partition function through the relation,
\begin{equation}\label{ener}
	\langle E \rangle = -\frac{\partial \ln Z}{\partial \beta}.
\end{equation}
At equilibrium temperature (corresponding to 
$\beta = \beta_0$), the average energy becomes
\begin{equation}\label{e-1}
	\langle E \rangle = -\frac{\partial \ln Z}{\partial \beta} \bigg|_{\beta = \beta_0} = \frac{-1}{Z} \frac{\partial Z}{\partial \beta} \bigg|_{\beta = \beta_0}.
\end{equation}
Similarly, the average of the squared energy can be written as,
\begin{equation}\label{e2}
	\langle E^2 \rangle = \frac{\partial^2 Z}{\partial \beta^2} \frac{1}{Z} \bigg|_{\beta = \beta_0}.
\end{equation}
The energy fluctuation, 
$\Delta E$, in the system is then given by

\begin{equation}
	(\Delta E)^2 = \langle E^2 \rangle - \langle E \rangle^2,
\end{equation}
which takes the form,
\begin{equation}\label{e-e}
	(\Delta E)^2 = \frac{\partial^2 Z}{\partial \beta^2} \frac{1}{Z} \bigg|_{\beta = \beta_0} - \left( \frac{-1}{Z} \frac{\partial Z}{\partial \beta} \bigg|_{\beta = \beta_0} \right)^2.
\end{equation}
The entropy and the partition function of the system are connected by the formula,
\begin{align}
	S(\beta) &= -\frac{\partial F}{\partial T}, \\
	&= k_B \ln Z(\beta) + k_B \beta \langle E \rangle,
\end{align}
where $F = -k_B T \ln Z$, is the Helmholtz free energy. Small energy fluctuations around thermal equilibrium are taken into account by differentiating $S(\beta)$ two times with respect to $\beta$ as shown below \cite {LandauStatMech}. 

\begin{equation}\label{entropy}
	S(\beta)'' = k_B \left( \frac{\partial^2 Z}{\partial \beta^2} \frac{1}{Z} - \left( \frac{-1}{Z} \frac{\partial Z}{\partial \beta} \right)^2 \right).
\end{equation}
This expression is similar to '$(\Delta E)^2$' , in equation (\ref{e-e}). Now, the fluctuations in energy can be expressed in terms of entropy as,
\begin{equation}\label{sigma}
	\langle \sigma^2 \rangle = \frac{S(\beta)''}{k_B}.
\end{equation}
This relation which governs the system's stability near equilibrium, enables the estimation of horizon energy fluctuations within different gravity frameworks, under the assumption that the universe behaves as a thermodynamic system.

\section{Fluctuations in horizon energy}
In this section, we investigate the statistical nature of spacetime by analysing thermal fluctuations in horizon energy, treating the cosmological horizon as a thermodynamic system. Since horizons possess temperature and entropy—both in black hole spacetimes and in cosmological settings—it is natural to associate an energy and corresponding fluctuations with them, as in any canonical ensemble \cite{Bekenstein1973,Hawking1975,Jacobson1995}. Energy fluctuations encode information about the underlying microscopic degrees of freedom and provide a sensitive probe of thermodynamic stability cite{Padmanabhan2010}. 

First, we analyse horizon energy fluctuations in (n+1)-dimensional Einstein gravity, which serves as the baseline theory with standard Bekenstein–Hawking entropy. We then extend the analysis to Gauss–Bonnet gravity and to more general Lovelock gravity, where higher-curvature corrections modify the entropy–area relation \cite{Cai2002,Wald1993}. We then examine horizon energy fluctuations in the presence of quantum-deformed entropy, motivated by quantum gravity considerations, where corrections to horizon entropy lead to measurable changes in fluctuation behaviour \cite{Padmanabhan2010,KrishnaMathew2019}. 
\subsection{Horizon energy fluctuation in (n+1) Einstein gravity}
 Here, we are going to calculate the horizon energy fluctuation in (n+1) Einstein gravity. In (n+1) Einstein gravity, the horizon entropy is given by  
\begin{equation}\label{entropy1}
	 S = \frac{n\Omega_{n}\tilde{r}_A^{n-1}}{4L_p^{n-1}}.
\end{equation}
The temperature of horizon is defined as 
\begin{equation*}
	T = \frac{\hbar}{2\pi k_B\tilde{r}_A} .
\end{equation*}
Differentiating equation (\ref{entropy1}), with respect to $\beta$, we get
\begin{equation}
	S' = \frac{\partial S}{\partial \tilde{r}_A}\frac{\partial  \tilde{r}_A}{\partial \beta}  = \frac{n(n-1)\Omega_{n}\tilde{r}_A^{n-2}}{4L_p^{n-1}}\frac{d\tilde{r}_A}{d\beta},
\end{equation}
which can be simplified as, 
\begin{equation}	
	S'= \frac{\hbar n(n-1)\Omega_{n}\tilde{r}_A^{n-2}}{8\pi L_p^{n-1}}.
\end{equation}
Differentiating the above equation once again with respect to $\beta$, we get,
\begin{equation}
	S'' = \frac{\hbar^2 n(n-1)(n-2)\Omega_{n}\tilde{r}_A^{n-3}}{16\pi^2 L_p^{n-1}}.
\end{equation}
Now the fluctuation in horizon energy can be expressed as,
\begin{equation}
	<\sigma^2> = \frac{S''}{k_B} =  \frac{\hbar^2 n(n-1)(n-2)\Omega_{n}\tilde{r}_A^{n-3}}{16\pi^2 k_B L_p^{n-1}}.
\end{equation}
Taking the derivative of fluctuation with respect to cosmic time, we get, 
\begin{equation}
	\frac{d}{dt}\Big(\frac{S''}{k_B}\Big) = \frac{\hbar^2 n(n-1)(n-2)(n-3)\Omega_{n}\tilde{r}_A^{n-4}}{16\pi^2 k_B L_p^{n-1}}\dot{\tilde{r}}_A.
\end{equation}
With the help of Friedmann and continuity equations, we can express $\dot{\tilde{r}}_A$ in terms of the equation of the state parameter as,  
\begin{equation}\label{rd}
\dot{\tilde{r}}_A = \frac{n}{2}H\tilde{r}_A(1+\omega).
\end{equation}
Following this, one can write,
\begin{equation}
		\frac{d}{dt}\Big(\frac{S''}{k_B}\Big) = \frac{\hbar^2 n(n-1)(n-2)(n-3)\Omega_{n}\tilde{r}_A^{n-4}}{16\pi^2 L_p^{n-1}}\frac{n}{2}H\tilde{r}_A(1+\omega).
\end{equation}
Thus at $\omega = -1$ ; 
\begin{equation*}
	\frac{d}{dt}\Big(\frac{S''}{k_B}\Big) = 0.
\end{equation*}
This shows that the fluctuation in horizon energy becomes a constant in the final de Sitter state when $\omega =-1$. This clearly demonstrates that the horizon energy fluctuations asymptotically attain a constant value when the universe evolves to the final de Sitter state.
\subsection{Horizon energy fluctuations in Gauss-Bonnet gravity}

Building upon the foundational studies in Einstein gravity, we extend the analysis on the horizon energy fluctuations to Gauss-Bonnet gravity, a higher-curvature generalization of Einstein gravity. Gauss-Bonnet gravity itself arises as a natural extension of Einstein's theory in higher-dimensional spacetime,  

In the context of Gauss-Bonnet gravity, the horizon entropy takes a more complex form given by,
\begin{equation}\label{eqn:entropy}
S=\frac{A}{4L^{n-1}_p}[1+\frac{n-1}{n-3}\frac{2 \tilde \alpha}{\tilde r^2_A}],
\end{equation}
where $A=n\Omega_n\tilde r^{-1}_A$, the area of the apparent horizon, $\ n \geq 4$, $\tilde\alpha =(n-2)(n-3)\alpha$, $\alpha$ being the Gauss-Bonnet coefficient which is positive. 

 Now, the partial derivative of horizon entropy with respect to ${\beta}$ can be found as,

\begin{equation}\label{}
S'=\frac{\hbar}{2\pi}\frac{n\Omega_n(n-1)}{4L^{n-1}_p}[\tilde r^{n-2}_A+2\alpha \tilde r^{n-4}_A].
\end{equation}
Differentiating the above equation once again with respect to ${\beta}$, we get 
\begin{equation}\label{eqn:sqare of partial}
S'' =(\frac{\hbar}{2\pi})^2\frac{n\Omega_n(n-1)}{4L^{n-1}_p}[(n-2)\tilde r^{n-3}_A+2\alpha(n-4)\tilde r^{n-5}_A].
\end{equation}
Taking the derivative of fluctuations with respect to cosmic time, we arrive at 
\begin{equation}\label{}
\frac{d}{dt}\Big(\frac{S''}{k_B}\Big)=(\frac{\hbar}{2\pi })^2\frac{n\Omega_n(n-1)}{4 k_B L^{n-1}_p}[(n-2)(n-3)\tilde r^{n-4}_A+2\alpha(n-4)(n-5)\tilde r^{n-6}_A]\dot{\tilde{r}}_A.
\end{equation}
With the help of Friedmann equation in Gauss-Bonnet gravity and continuity equation, one can express $\dot{\tilde{r}}_A$ as,
 
\begin{equation}\label{eqn:gbraddot}
\dot{ \tilde{r}}_A =\frac{8\pi L^{n-1}_p}{(n-1)}{{\tilde{r}_A^{3}H} { (1+\omega)\rho}\over (1+{2\tilde\alpha \tilde{r}_A^{-2}})},
\end{equation}
where $\omega$ is the equation of state parameter. When the Universe attains its final de-sitter state, $\omega=-1$ and $\dot{ \tilde{r}}_A =0$. Then one can easily arrive at,

\begin{equation}\label{eqn:final}
{d\over dt}(\frac{\partial^2S}{\partial\beta^2})= 0.
\end{equation}
Here, it is shown that the fluctuation in horizon energy becomes a constant in the final de Sitter state of the universe when $\omega=-1$. This result supports the interpretation of the de Sitter universe as a thermodynamically stable state, in the context of Gauss-Bonnet gravity.
\subsection{Horizon energy fluctuations in Lovelock Gravity}
We will now move to Lovelock gravity which is a generalization of Gauss-Bonnet gravity, such that the Lagrangian consists of dimensionally extended Euler densities. The horizon entropy in Lovelock gravity can be expressed as,
\begin{equation}\label{eqn:loveentro}
S= \frac{A}{4L^{n-1}_p} \sum_{i=1}^m \frac{i(n-1)}{(n-2i+1)} \hat{c_i}{\tilde r_A}^{2-2i}
\end{equation}
where $A=n\Omega_n \tilde r^{-1}_A$.
Taking the derivative of Eq. (\ref{eqn:loveentro}) with respect to the $\beta$, we get

\begin{equation}\label{}
\begin{aligned}
 S'= &(\frac{\hbar}{2\pi})^2\frac{n\Omega_n(n-1)}{4L^{n-1}_p}[ \tilde r^{n-1}_A][(n-1)\sum_{i=1}^m\frac{i\tilde{c_i}{(2-2i)}{\tilde r^{-2i}_A}}{(n-2i+1)}+(n-2)(n-1)\sum_{i=1}^m\frac{i\tilde{c_i} \tilde r^{-2i}_A}{(n-2i+1)} \\
  &\quad 
+(n-1)\sum_{i=1}^m\frac{i\tilde{c_i}{(2-2i)} \tilde r^{-2i}_A}{(n-2i+1)}+\sum_{i=1}^m\frac{i\tilde{c_i}{(2-2i)(2-2i+1)}{\tilde r^{-2i}_A}}{(n-2i+1)}]
\end{aligned}
\end{equation}
Differentiating the above equation once again with respect to $\beta$, we get
\begin{equation}\label{}
\begin{aligned}
S''=& (\frac{\hbar}{2\pi})^2\frac{n\Omega_n(n-1)}{4L^{n-1}_p}[\tilde r^{n-1}_A]\\
  &\quad\sum_{i=1}^m\frac{i\tilde{c_i}{\tilde r^{-2i}_A}}{(n-2i+1)}[(n-1)(2-2i)+(n-2)(n-1)+(n-1)(2-2i)+(2-2i)(2-2i+1)]
  \end{aligned}
\end{equation}

Differentiating the above equation with respect to cosmic time, we arrive at, 
\begin{equation}\label{K}
\begin{aligned}
\frac{d}{dt}\Big(\frac{S''}{k_B}\Big) = &(\frac{\hbar}{2\pi})^2\frac{n\Omega_n(n-1)}{4 k_B L^{n-1}_p}[\tilde r^{n-1}_A(\sum_{i=1}^m K_i\frac{i(\tilde{c_i})({-2i})}{(n-2i+1)}(\tilde r^{-2i-1}_A)(\dot \tilde { r_A}))\\
  &\quad+(\sum_{i=1}^m K_i \frac{i(\tilde{c_i})({\tilde r^{-2i}_A})}{(n-2i+1)}(n-1)(\tilde r^{n-2}_A)(\dot \tilde {r_A})].
    \end{aligned}
\end{equation}
Here, $K_i= (n-1)(2-2i)+(n-2)(n-1)+(n-1)(2-2i)+(2-2i)(2-2i+1) $ (for convenience). With the help of Friedmann equation in Lovelock gravity and continuity equation, we can express $\dot{\tilde{r}}_A$ in terms of the equation of the state parameter as,
\begin{equation}
\dot \tilde{r_A}=\frac{16\pi L^{n-1}_p H(1+\omega)\rho}{\sum_{i=1}^m 2ic_i r^{-2i-1}_A}.
\end{equation}
Now, when the universe approaches the final de-sitter state, $\omega=-1$ and $\dot{ \tilde{r}}_A =0$. Then from equation (\ref{K}, we arrive at, 
\begin{equation}
\frac{d}{dt}(\frac{\partial^2 S}{\partial\beta^2})=0.
\end{equation}
This implies that the fluctuation in horizon energy attains a constant value in the final de-sitter state of the universe, just like an ordinary thermodynamic system in the context of Lovelock gravity.
\subsection {Horizon energy fluctuations in the framework of quantum deformed entropy}
In classical general relativity, the entropy associated with the horizon of a black hole or the apparent horizon of the universe follows the Bekenstein–Hawking area law, $S=A/G$, where $A$ is the horizon area and $G$ is Newton’s gravitational constant. At quantum scales, spacetime is expected to possess a discrete or quantized structure. This discreteness deforms the standard entropy–area relation, leading to a quantum-deformed (or q-deformed) entropy given by\cite{Bekenstein1973,Hawking1975,Chen2024,Das2002,Carlip2000,CaiCaoHu2009,ZhuRen2009,Medved2004},

\begin{equation}\label{qentro}
S=\frac{\pi(sin(\frac{\lambda\tilde r_A^2}{G}))}{sin\lambda}.
\end{equation}
The temperature associated with the apparent horizon is given by
\begin{equation}\label{}
T=\frac{1}{2\pi \tilde r_A}. 
\end{equation}
Taking the derivative of Eq.(\ref{qentro}) with respect to $\beta$, one gets,
\begin{equation}\label{}
S'=
{\frac{\lambda k_B}{G sin\lambda}}{cos(\frac{\lambda{{\tilde r^2_A}}}{G})}\tilde r_A.
\end{equation}
Differentiating the above equation once again with respect to $\beta$, we get
\begin{equation}\label{}
S''=({\frac{\lambda k_B^2}{2\pi G sin\lambda})}[cos(\frac{\lambda{\tilde r^2_A}}{G})-{2 \tilde r^2_A}{sin({\frac{\lambda{{\tilde r^2_A}}}{G}})}].
\end{equation}
Taking the derivative of the above equation with respect to cosmic time, we arrive at
\begin{equation}\label{}
\frac{d}{dt}\Big(\frac{S''}{k_B}\Big)={-\frac{\lambda k_B^2 }{\pi G sin\lambda}}[{\frac{\lambda}{G}}sin(\frac{\lambda{\tilde r^2_A}}{G})+{2}sin(\frac{\lambda{\tilde r^2_A}}{G})+2{\tilde r^2_A} \frac{\lambda}{G} cos(\frac{\lambda{\tilde r^2_A}}{G})] {\tilde r_A}{\dot{\tilde{r}}_A}.
\end{equation}
Now, we can express $\dot{\tilde{r}}_A$ in terms of the equation of the state parameter as \cite{Chen2024}, 
\begin{equation}\label{}
\dot{\tilde{r}}_A={\frac{4 \pi G sin (\lambda)}{\lambda}}{\frac{\tilde{r}_A^3 H (1+\omega)\rho}{\cos ({\frac{\lambda \tilde{r}_A^2}{G}})}}.
\end{equation}
When the universe attains the final de-sitter state, $\omega= -1$ and $\dot{\tilde{r}}_A$ vanishes. Hence in the final de-sitter epoch, one can write,
\begin{equation}\label{}
\frac{d}{dt}(\frac{\partial^2 S}{\partial\beta^2}) =0.
\end{equation}

Thus, in the final de Sitter state, when $\omega= -1$, the horizon energy fluctuations attains a constant value, resembling that of a conventional thermodynamic system governed by q-deformed entropy.
\section{Horizon energy fluctuations, horizon entropy maximization and the holographic equipartition- The possible connection}

Our analysis of horizon energy fluctuations reveals a particularly significant and universal behavior in the
asymptotic de Sitter limit. We find that as the equation-of-state parameter approaches $\omega = -1$, corresponding
to a de Sitter phase, the horizon energy fluctuations attains a constant value. Remarkably, this
stabilization is independent of the underlying gravitational dynamics and persists across Einstein gravity,
Gauss-Bonnet gravity, Lovelock gravity, as well as in the presence of quantum-deformed entropy corrections. In all the gravity models considered, the dynamical evolution of fluctuations slows down as $\omega \to -1$, eventually freezing to
a constant value. Such a behavior strongly suggests that the cosmological horizon reaches a thermodynamically stable configuration in the de Sitter phase. From a statistical perspective, the horizon behaves analogously to a canonical ensemble that has reached thermal equilibrium, reinforcing the interpretation of the de Sitter universe as a final equilibrium state of cosmic
evolution. 

Earlier investigations within the emergent gravity paradigm provide an important complementary viewpoint.
According to the law of emergence, the expansion of cosmic space is driven by the difference between the number
of degrees of freedom on the horizon surface and those residing in the bulk. It has been shown that when the
equation-of-state parameter approaches $\omega = -1$, this difference vanishes, leading to the holographic equipartition condition
$N_{\text{surf}} = N_{\text{bulk}}$ \cite{Padmanabhan2012,Krishna2024}. 

Within this framework, horizon entropy plays a central role. Previous studies have demonstrated that when
$N_{\text{surf}} = N_{\text{bulk}}$, the horizon entropy reaches a maximum constant value, analogous to the
maximum entropy principle in conventional thermodynamics. This entropy maximization occurs universally in Einstein, Gauss-Bonnet, and Lovelock gravity theories, despite the modified entropy-area relations arising
from higher-curvature terms. The de Sitter phase thus emerges as the configuration in which the horizon entropy is maximized, consistent with the second law of thermodynamics applied to spacetime horizons.

Our results add a new and independent thermodynamic element to this picture by demonstrating that horizon
energy fluctuations also stabilize precisely in the same limit, $\omega \to -1$. The coincidence of entropy
maximization, holographic equipartition, and fluctuation stabilization is highly nontrivial and suggests that
these phenomena are deeply interconnected. In ordinary thermodynamic systems, equilibrium is characterized
simultaneously by maximum entropy and stationary fluctuations. The present analysis shows that an analogous
structure exists for cosmological horizons, strengthening the analogy between spacetime dynamics and
macroscopic thermodynamic systems.

Taken together, these results point toward a coherent and unified thermodynamic interpretation of cosmic
evolution. As the universe evolves toward the de Sitter phase, it approaches a state where the horizon entropy
is maximized, the surface and bulk degrees of freedom are balanced, and energy fluctuations become constant.
These features collectively identify the asymptotic de Sitter universe as the natural thermodynamic end state
of the universe. The stabilization of horizon energy fluctuations thus provides a novel and robust
thermodynamic signature of holographic equipartition and supports the broader view that gravity and cosmic
expansion arise from underlying statistical and thermodynamic principles.

\section {Conclusion}

In this work, we have investigated thermal energy fluctuations associated with cosmological horizons by treating the horizon as a thermodynamic system within the canonical ensemble framework. By computing the second derivative of entropy with respect to the inverse temperature, we quantified horizon energy fluctuations and used them as a diagnostic of thermodynamic stability. Our analysis was carried out systematically across Einstein gravity, Gauss-Bonnet gravity, Lovelock gravity, and scenarios incorporating quantum-deformed entropy. This unified approach allowed us to examine how higher-curvature corrections and quantum modifications to the entropy-area relation influence the statistical behavior of horizon energy fluctuations. The results demonstrate that, despite differences in the underlying gravitational dynamics, the qualitative thermodynamic behavior remains robust across these theories.

A central outcome of our analysis is the identification of the de Sitter phase, characterized by the equation of state parameter $\omega = -1$, as an equilibrium configuration. In this asymptotic limit, the horizon energy fluctuations stabilize to a constant value, indicating the attainment of a thermodynamic equilibrium-like state. This behavior mirrors that of ordinary macroscopic systems in canonical ensembles, where energy fluctuations cease to evolve once equilibrium is reached. Importantly, this stabilization persists not only in Einstein gravity but also in Gauss-Bonnet and Lovelock gravities, as well as in the presence of quantum-deformed entropy corrections. The result therefore provides strong evidence that the late-time de Sitter universe represents a thermodynamically preferred state, independent of the specific gravitational theory under consideration.

Our findings acquire further significance when interpreted within the emergent gravity paradigm. Previous studies have shown that the law of emergence implies holographic equipartition, $N_{\text{surf}} = N_{\text{bulk}}$, precisely when $\omega \to -1$ \cite{Padmanabhan2012,Komatsu2022}. This condition is also leads to the maximization of horizon entropy, signaling the end of cosmic evolution toward thermodynamic equilibrium. In the present work, we have shown that horizon energy fluctuations likewise become constant in this limit, thereby completing a consistent thermodynamic picture: when $\omega \to -1$, the universe simultaneously attains holographic equipartition, maximum horizon entropy, and stabilized energy fluctuations. These three features together strongly support the interpretation of the asymptotic de Sitter state as the final equilibrium configuration of the universe.

More broadly, our results reinforce the view that cosmic expansion and gravitational dynamics may be understood as emergent phenomena governed by statistical and thermodynamic principles. Energy fluctuations provide a sensitive probe of the microscopic degrees of freedom underlying spacetime and offer a complementary perspective to entropy-based arguments. The persistence of fluctuation stabilization across higher-curvature and quantum-corrected gravity theories suggests a universal thermodynamic signature of spacetime evolution. We expect that extending this framework to nonequilibrium settings, alternative ensembles, or dynamical dark energy models may yield further insights into the statistical origin of gravity and the thermodynamic nature of cosmic acceleration.

\end{document}